# The Structure of GaSb Digitally Doped with Mn

G. I. Boishin[a,†], J. M. Sullivan[a,b,†,*] and L. J. Whitman[a,*]
[a]*Naval Research Laboratory, Washington, DC 20375*
[b]*Northwestern University, Evanston, IL 20208*

Cross sectional scanning tunneling microscopy (XSTM) and density functional theory have been used to characterize the structure of GaSb digitally doped with Mn. The Mn dopants are found in both isolated substitutional form as well as in large clusters of zinc-blende MnSb commensurate with the surrounding GaSb matrix. Theoretical calculations predict that these two forms of Mn in the digitally doped layers will have a very different appearance in XSTM images. Substitutional Mn enhances the local density of states near the surface, thus appearing higher in a filled-state image. In contrast, MnSb clusters induce substantial structural relaxation at the {110} surface, and therefore appear as localized depressed regions with negligible perturbation of the surrounding GaSb.

## Introduction

Ferromagnetic materials that can be easily integrated into semiconductor electronics are crucial for so-called "spintronic" applications. Concomitant with the obvious structural requirements—for example, ferromagnetic materials lattice matched to GaAs and Si—is the desire for room temperature operation. Diluted magnetic semiconductors, such as Mn-doped GaAs,[1,2] are designed to fulfill the first of these requirements, whereas fulfilling the requirement of room-temperature operation has been more difficult. Recently some degree of success in raising the ferromagnetic ordering temperature has been obtained with digitally doped samples.[3,4] However, there has been little progress in understanding the atomic-scale structure of the doping profile in such samples.

In this Brief Report, we combine cross-sectional scanning tunneling microscopy and density functional theory to characterize the structure and distribution of dopants in GaSb/Mn digital alloys. High-resolution atomic-scale images of the ($\bar{1}$10) cross- sectional surface show a number of characteristic localized defects. These localized defects are analyzed using simulations of the XSTM experiments derived from density functional calculations and interpreted as isolated substitutional Mn dopants on the Ga sublattice, and clusters of zinc-blende MnSb commensurate with the GaSb matrix. To the best of our knowledge, this is the first report of a combined theoretical-experimental effort to identify zinc-blende MnSb clusters in GaSb.

## Experimental

The sample was grown at the University at Buffalo and its magneto-electronic properties have been reported previously.[3] Briefly, it was grown on a *p*-type GaAs (001) substrate with a thick, fully relaxed GaSb buffer layer. The GaSb/Mn digital alloy, grown at 275 °C, consists of ½ monolayer of Mn deposited every 16 ML.[3] The sample was first thinned to ~200 μm, and then loaded into the chamber, scribed, and cleaved *in situ* to expose a ($\bar{1}$10) surface.[5] The XSTM measurements were performed in constant-current mode in an ultrahigh-vacuum chamber with base pressure <$1 \times 10^{-10}$ Torr. All images shown here are of filled electronic states, nominally showing the topography of the Sb sublattice on the cross-sectional surface.

Figure 1(a) shows a nanoscale overview of the sample, including the interface between the buffer layer and the digitally doped region; the first 18 periods of the GaSb/Mn digital alloy with a period of approximately 5 nm are indicated. Three different terrace levels are observed as well as numerous steps. (The high step density on the cleavage surface is presumably a result of the dislocation network caused by the 8% lattice mismatch between the fully-relaxed GaSb buffer layer and the GaAs substrate.) The GaSb buffer layer displays a smooth, uniform composition; in comparison, filled-state images of the digitally doped region show frequent nanometer-scale depressions (dark areas) aligned along the nominal Mn-containing layers. Figure 1(b) shows a higher resolution image of the region near the interface. Note that XSTM of the ($\bar{1}$10) surface reveals only every other row in the [001] growth direction, so that the observed period of the depressions is 8 Sb rows, consistent with the intended Mn doping profile (every 16 growth layers).

The typical dimensions of the depressions are about 1 to 4 nm along [110] (in the plane of the deposited layer), and 1 to 3 rows (0.6 to 2 nm) in the [001] growth direction. There is little apparent perturbation of the lattice surrounding these features, and no evidence of local band bending. Recent first principles calculations found that zinc-blende MnSb is lattice-matched to GaSb with an optimal lattice constant of 6.2 Å, but is mechanically unstable.[6,7] However, the theoretical results also predict that a small amount of Mn can be incorporated in GaSb.[6] Considering that our sample has been digitally doped with only ½ monolayer, we suggest that the depressions observed in the XSTM images are zinc-blende MnSb clusters. This is consistent with the fact that the features appear to have the same lattice structure as the host lattice, rather than that of pure Mn or MnSb in the NiAs structure. The 0.1 nm depth of the depressions could result from reduced charge transfer from Mn to Sb, post-cleaving atomic relaxation, or combination of such effects.

In addition to these localized depressions we observe a number of other characteristic features; Fig. 2 shows an atomic resolution XSTM image of the Mn digitally doped region. In addition to the depressions there are several atomic-size bright and dark features not normally observed in GaSb,[5,8] suggesting the presence of additional foreign atoms or defects. The small dark features, which are present in both the GaSb buffer layer and digitally doped region, are characteristic of As substitutional defects[9] (substituting for Sb) and are believed to be present because the As cell used to prepare the GaAs substrate was still hot during the GaSb growth. As we discuss later in the paper, the bright features, which correlate strongly with the period of the Mn doping, are most likely due to randomly substituted Mn dopants on the Ga sublattice.

**Theoretical**

To help interpret our experimental results we performed electronic structure calculations in the generalized gradient approximation of Perdew and Wang,[10] followed by simulations of constant-current XSTM in the approach of Tersoff and Hamann.[11] A plane-wave basis set was used in the projector augmented wavefunction method[12] as implemented in the VASP code to self-consistently determine the ground state charge density and Kohn-Sham wavefunctions, with total energies converged to 0.1 meV.[13] The wavefunctions were calculated for a single wavevector in the Brilloun zone of the slab supercells at a kinetic energy cutoff of 270 eV. For all the calculations, we used as the lattice parameter our theoretically optimized value of 6.2 Å. For the simulations of filled-state XSTM images, the local density of states (LDOS) was integrated over an energy range of 2 eV up to the Fermi level.

We modeled the GaSb ($\bar{1}10$) cleavage surface in a supercell slab geometry consisting of 5 layers of GaSb separated by a vacuum region of 14 Å; the bottom layer of the slab was passivated to simulate semi-infinite bulk material. We considered isolated Mn substitutional defects in the 1st and 2nd layers near the surface and MnSb clusters. For the cases of isolated Mn dopants we used a 4×4 surface unit cell; for simulating clusters of MnSb in GaSb, we used a 7×1 surface unit cell with 4 bilayers of MnSb in the [001] growth direction. To minimize spurious superlattice effects, for supercells containing a single substitutional Mn, only atoms within 7.0 Å of the Mn site were allowed to relax during the minimization of the energy. For the supercell with 4 MnSb bilayers, all layers but the bottom passivating layer were relaxed during the energy minimization. The criterion for convergence of the structural optimizations was that the energy difference between two structural configurations be less than 1 meV.

Figures 3(a) and 3(b) show the simulated filled-state XSTM images for Mn substitutional dopant atoms in the 1$^{st}$ and 2$^{nd}$ layers, respectively. Both of these dopants have a LDOS near Mn that is slightly larger than the unperturbed surface LDOS, i.e. an increased tip-sample separation. Specifically, these Mn substitutionals increase the LDOS on surface Sb atoms to which they are bonded. Thus Mn substitutional in the 1$^{st}$ layer is associated with enhanced LDOS on two Sb atoms [below the Mn site in Fig. 3(a)] and Mn substitutional in the 2$^{nd}$ layer is signified by an enhanced LDOS on a single Sb surface atom [above the Mn site in Fig. 3(b)]. Overall the simulated spectra are similar to the bright features observed in Fig. 2 and are consistent with those of substitutional Mn dopants in GaAs.[14,15] Thus we propose that the atomic-scale bright protrusions observed in Fig. 2 are associated with isolated Mn dopants, and that the larger hillocks result from overlap of the LDOS of substitutional Mn defects in close proximity.

To determine what effects clusters of MnSb commensurate with the GaSb lattice would have, we calculated the geometric and electronic structure of 4 bilayers of MnSb in GaSb. A structural model of the optimized supercell is shown in Fig. 4; note that for clarity, a 7×3 surface unit cell is shown. Although bulk MnSb is well lattice matched to bulk GaSb, there is substantial structural relaxation near the surface in the MnSb bilayers. The Mn atoms at the surface relax all the way into interstitial locations *below* the surface to form complexes with Mn in the subsurface layer. There is an accompanying ~1 Å relaxation of the Sb surface atoms into the surface in these bilayers. However, there is negligible relaxation of the surrounding GaSb. Thus small clusters of MnSb in GaSb should induce significant surface relaxation on the cleaved {110} surface but little perturbation to the surrounding GaSb.

The effects of these structural relaxations are evident in Fig. 5, where the simulated XSTM image of the four MnSb bilayers is displayed along with a comparison of theoretical and experimental line profiles. The two line profiles compare favorably, showing an approximate depth of 1 Å, which agrees well with the predicted structural parameters. In addition, both the theoretical and experimental line profiles exhibit similar structure in the bottom of the depression; the theoretical results indicate that this structure originates from the surface Mn which relaxes to subsurface locations.

**Summary**

Digitally doped GaSb/Mn alloys have been investigated with XSTM on the GaSb ($\bar{1}10$) surface and interpreted with electronic structure calculations. The atomic scale characterization reveals coexistence of isolated Mn substitutional dopants on the Ga sublattice and quasi two-dimensional islands of zinc-blende MnSb in the digitally doped layers. Atomic relaxation of the ($\bar{1}10$) cleavage surface on the MnSb islands produces Mn substitutional-interstitial complexes with distinct signatures in the XSTM images.

**Acknowledgements**

This work was supported by the Office of Naval Research and the Defense Advanced Research Projects Agency. Computational work was supported in part by a grant of HPC time from the DoD Major Shared Resource Center ASCWP. The sample used in this study was generously provided by G. B. Kim, M. Cheon, S. Wang and H. Luo at the University at Buffalo with support from the DARPA SPINS program through the Office of Naval Research under Grant N00014-00-1-0951.

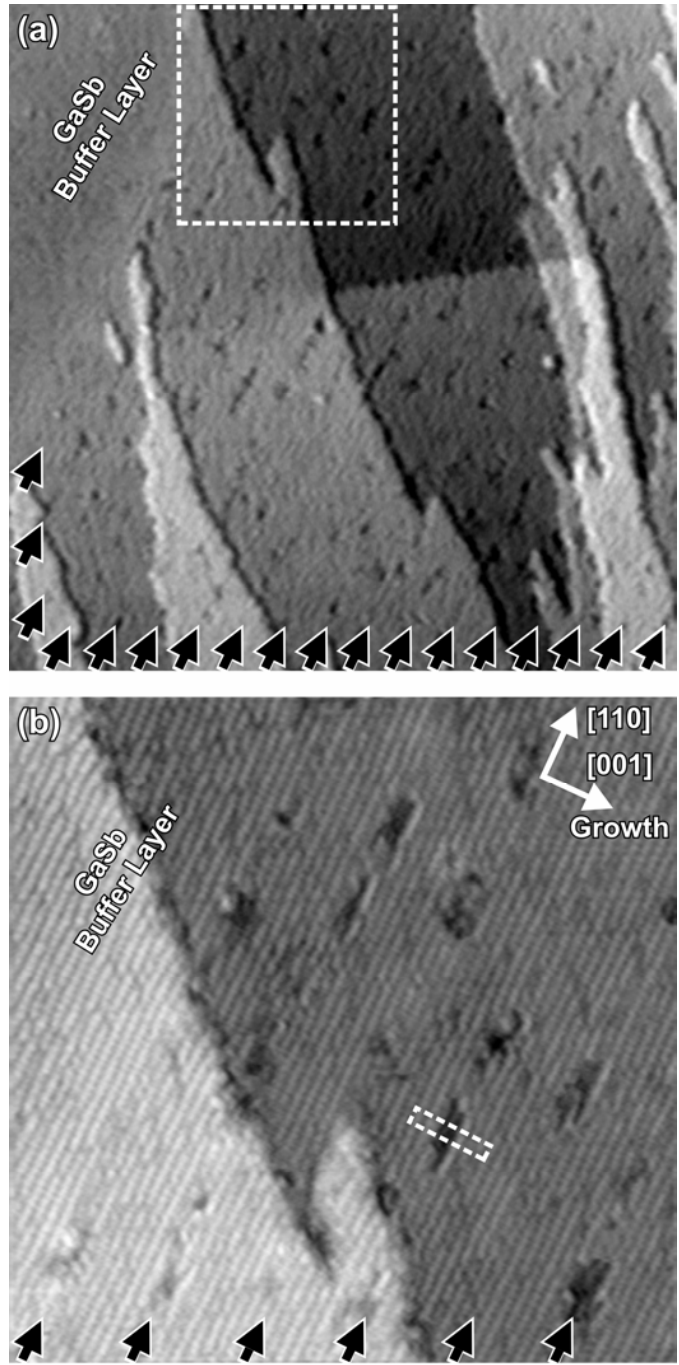

Fig. 1 Filled-state XSTM images of a GaSb/Mn digital alloy (2.30 V, 30 pA). (a) 87 nm × 87 nm area including the end of the GaSb buffer and the first 18 superlattice periods, (b) 35 nm × 35 nm area indicated in (a) by the dashed square. Nominal positions of the Mn layers are indicated by the arrows. In (b) the dashed lines indicate the region for the line profile in Fig. 5 (see text).

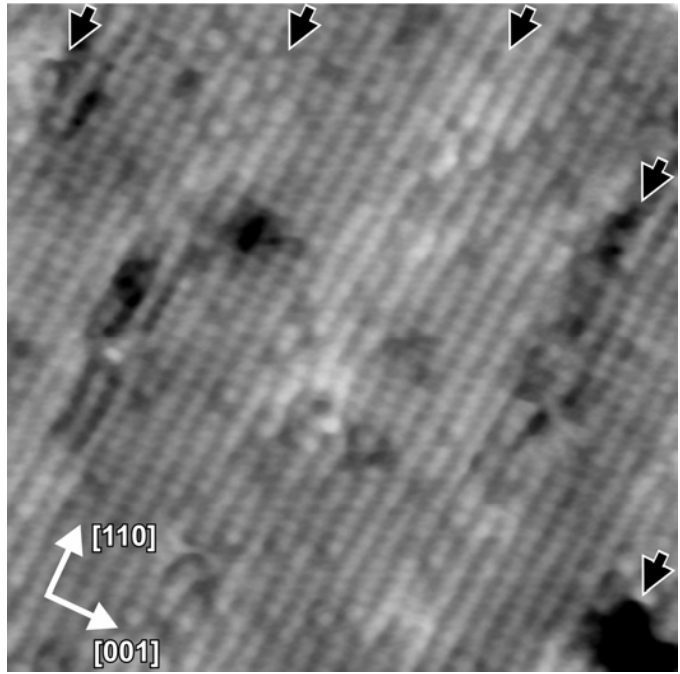

Fig. 2 Filled-state XSTM image of GaSb digitally doped with Mn, 18 nm × 18 nm (1.90 V, 40 pA). Arrows indicate the nominal positions of the Mn layers.

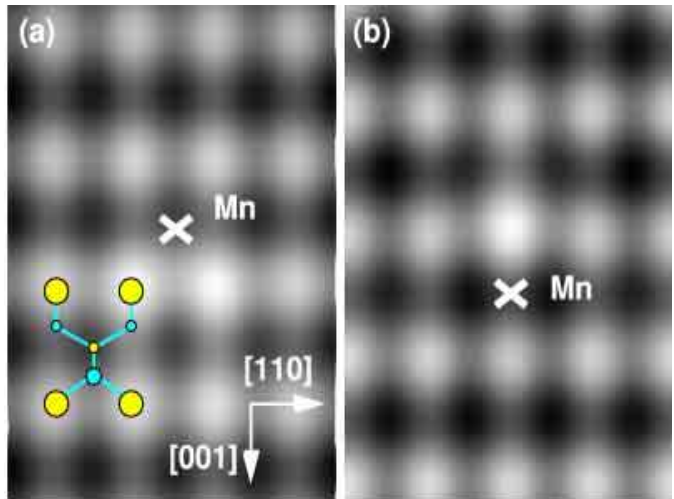

Fig. 3 Simulated filled-state XSTM images of Mn substitutional dopants near the GaSb (110) surface. Ga and Sb atoms are shown in cyan (gray) and yellow (light gray) respectively. (a) Mn substitutional dopant in the 1st layer, (b) Mn substitutional dopant in the 2nd layer. A structural model of the underlying lattice is shown in (a) for the first two layers. The location of the Mn substitutional is indicated in each image.

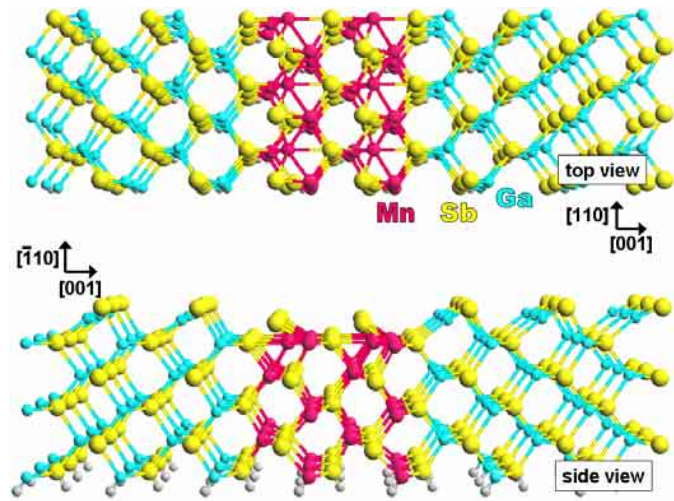

Fig. 4 (Color online) Top view (upper) and side view (lower) structural model showing optimized coordinates of 4 bilayers of MnSb in GaSb. Ga, Sb and Mn atoms are shown in cyan (gray), yellow (light gray) and red (dark gray) respectively. For clarity a 7×3 surface unit cell is shown.

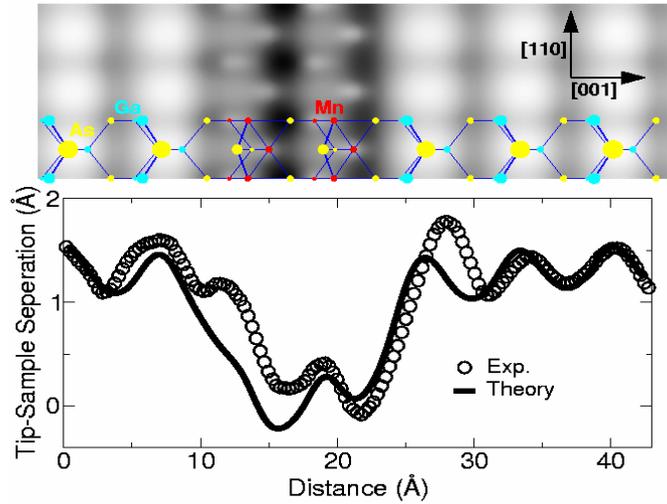

Fig. 5 (Color online) Simulated filled-state XSTM results for 4 bilayers of MnSb in GaSb. Ga, Sb and Mn atoms are shown in cyan (gray), yellow (light gray) and red (dark gray) respectively. (a) simulated STM image for the (110) surface with the top few atomic layers projected onto the data for clarity; (b) comparison of experimental and theoretical line profiles through the MnSb bilayers. In (b) the zero of the vertical scale is arbitrary. The experimental line profile is from the dashed region indicated in Fig. 1(b).

\* Corresponding authors: whitman@nrl.navy.mil, james.sullivan@nrl.navy.mil

†Also at: Nova Research Inc., Alexandria, VA 22308